\documentclass[
 amsmath,amssymb,
 aps,
]{revtex4-2}

\usepackage{graphicx} 
\usepackage[utf8x]{inputenc}
\usepackage[T1]{fontenc}
\usepackage{amsmath}
\usepackage{amssymb}
\usepackage{xcolor}
\usepackage{subcaption}
\usepackage{braket}

\usepackage[pdftex,linkcolor=black,pdfborder={0 0 0}]{hyperref} 
\usepackage{calc} 
\usepackage{enumitem} 

\usepackage{dcolumn}
\usepackage{bm}

\newcommand{\ep}{\epsilon}
\newcommand{\m}{\mathcal}
\newcommand{\al}{\alpha}
\newcommand{\bt}{\beta}
\newcommand{\ti}{\tilde}
\newcommand{\la}{\lambda}
\newcommand{\w}{\omega}
\newcommand{\p}{\partial}
\newcommand{\Ai}{\text{Ai}}

\begin{document}

\preprint{APS/123-QED}

\title{Prediction of Large Events in Directed Sandpiles}

\author{Dhruv Shah}

 \email{dhruvs04@mit.edu}
\affiliation{ Department of Physics, Massachusetts Institute of Technology MA 02139, USA }
 \affiliation{ International Centre for Theoretical Sciences, Tata Institute of Fundamental Research, Bangalore 560089, India}

\date{\today}

\begin{abstract}
The degree of predictability of large avalanche events in the directed sandpile model is studied. A waiting time based prediction strategy which exploits the local anticorrelation of large events is discussed. With this strategy we show analytically and numerically that large events are predictable to some extent, and that this predictability persists in the thermodynamic limit. We introduce another strategy which predicts large avalanches in the future based on the present excess density in the sandpile. We obtain the exact conditional probabilities for large events given an excess density, and use this to determine the exact form of the ROC predictability curves. We show that for this strategy, the model is predictable only for finite lattice sizes, and unpredictable in the thermodynamic limit. This behaviour is to be contrasted with previously established numerical studies carried out for Manna sandpiles.

\end{abstract}

\maketitle

\section{Introduction}

Self organized critical (SOC) systems \cite{bakbook, socreview} have been extensively studied over the past forty years. The criticality of such systems is not based on fine tuning of the system parameters. These systems are slowly driven, and in the critical state result in dissipation, the magnitude of which is power law distributed and scale free. SOC models, therefore, have been appropriate systems for the study of extreme events. An important question which arises is with regard to what degree these extreme events in SOC models can be predicted using past data. \\

Natural systems such as earthquakes \cite{bakbook, richter, saichev}, wildfires \cite{wildfire}, solar flares \cite{bakbook} show long-range spatial correlations and power-law behavior. These characteristic properties are well reproduced in simple SOC lattice models; this has led to a new paradigm in the prediction of extreme events in natural systems \cite{quake1}. Of the SOC lattice models, sandpile models are the most convenient to deal with \cite{revpile}. Multiple studies have used sandpiles and other SOC models to discuss the issue of predictability with regard to natural systems \cite{springblocksoc, quaketopile,  quaketopile2}. In \cite{springblocksoc}, a nonconservative SOC lattice model is shown to be equivalent to the well established Burridge Knockoff model for faults in earthquakes \cite{springblockog, springblock1d}. In \cite{quaketopile}, the authors identify precursors to large events in a variety of SOC models, analogous to seismic signatures before large earthquakes; and use these to plot prediction success rate curves. In \cite{quaketopile2}, a new sandpile-like automaton model is introduced which successfully reproduces the Gutenberg Richter law for energy distribution in earthquakes, as well as other empirical data of earthquakes from seismogenic regions.\\

The notion of predictability in sandpiles varies amongst different works. In \cite{montoya}, the sandpile is said to be predictable if, given any initial state of the sandpile, there exists an algorithm to determine the final recurrent state, such that the algorithm is faster than direct time evolution of the system. In this sense, they conjecture that the sandpile model is not predictable. A different kind of problem is short term prediction. Given a past time series of occurence of extreme events, one tries to predict whether an extreme event is likely in the next few timesteps \cite{shapoval, garber}. We will use the term `predictability' in this second sense.  \\

Next timestep prediction is based on a decision variable constructed out of present or past values of observables. In \cite{shapoval}, the decision variable is the fraction of almost-unstable sites on the sandpile. In \cite{shapoval2}, the decision variable is constructed with a weighted sum of past avalanche sizes, where the kernel is optimized for prediction efficiency. In both studies, they find that in the thermodynamic limit, sufficiently `large' events (to be defined later) in Manna sandpiles \cite{manna} can be predicted with a good success rate. However, BTW sandpiles do not seem to be predictable. A similar strategy is used in \cite{garber}, where subsequent predictions for extreme events are decreased if they occured recently in the past. This predictability, based on anticorrelation of extreme events, is based on a finite size effect, and vanishes in the thermodynamic limit. Both these strategies depend on continuous internal parameters. As a `measure' of the quality of prediction, predictability curves are plotted for a wide range of these parameters. \\

We will discuss here the prototypical Directed sandpile model \cite{revpile, directed}, because it is more tractable than the BTW or Manna sandpiles. This allows us to analytically calculate predictability curves (defined in detail later) and scaling forms in many cases. One can make many different kinds of `predictions' \cite{gneiting, scoringrules}, including predictions about the strength of the event, or given past data, provide the full a posteriori probability distribution of event sizes. We consider the simplest setting, where the prediction is a binary variable, in the form of an alarm, which is turned on if the extreme event is likely, and off otherwise. Of course, the strategy to turn the alarm on or off at a time $t+1$, takes the past data for times $(1,..,t)$ as input. We compare the predictability of two different strategies. The first is a waiting time strategy, which exploits the temporal anti-correlation of large avalanches. For an $L \times M$ directed sandpile, with $M \sim L^{1/2}$, we find that large avalanches are predictable to some extent. More importantly, contrary to \cite{garber, shapoval2}, this predictability is not a finite size effect and persists in the thermodynamic limit. Our second strategy makes use of measuring the excess density in the model, similar to \cite{shapoval}. We find that with such a strategy, avalanches are somewhat predictable for finite $L$, but predictability decreases with $L$ and vanishes in the thermodynamic limit. \\

This paper is structured as follows. In Sec. \ref{model}, we introduce the directed sandpile model, and general terminology related to the prediction problem. In Sec. \ref{waitingstrategy} we first discuss the efficiency of the waiting time strategy to predict large events. In Sec. \ref{globalevent} we show, using MC simulations that predicting rare events everywhere globally in the lattice is not possible under this strategy. In Sec. \ref{localevent} we obtain the waiting time distributions for events occurring in a local detection window. We present analytic expressions for the average waiting times. We plot predictability curves (defined later) using Monte Carlo data, and theoretically find expressions  for the prediction efficiency as a function of event size and system size, in the large $L$ limit. We show that there is a finite non zero gain in prediction efficiency in the thermodynamic limit. Next, we consider a different kind of prediction strategy based on the excess density in the sandpile. In Sec. \ref{xsheight} we use the known scaling properties of large avalanche events to exactly determine the predictability curves for the prediction strategy. We explicitly show that predictability diminishes with lattice size, and vanishes in the thermodynamic limit, and verify this using MC simulations. In Sec. \ref{xssize}, we first derive the conditional probabilities for size events, given an excess density, and use this to get exact expressions for the prediction curves. Again, we show that some non zero prediction advantage exists only for finite lattice size. This is also verified by MC simulations. Finally, in Sec. \ref{compare}, we compare these results with previously established numerical studies of other sandpile models.

\section{Model}\label{model}
\begin{figure}
    \centering
    \includegraphics[width=0.4\linewidth]{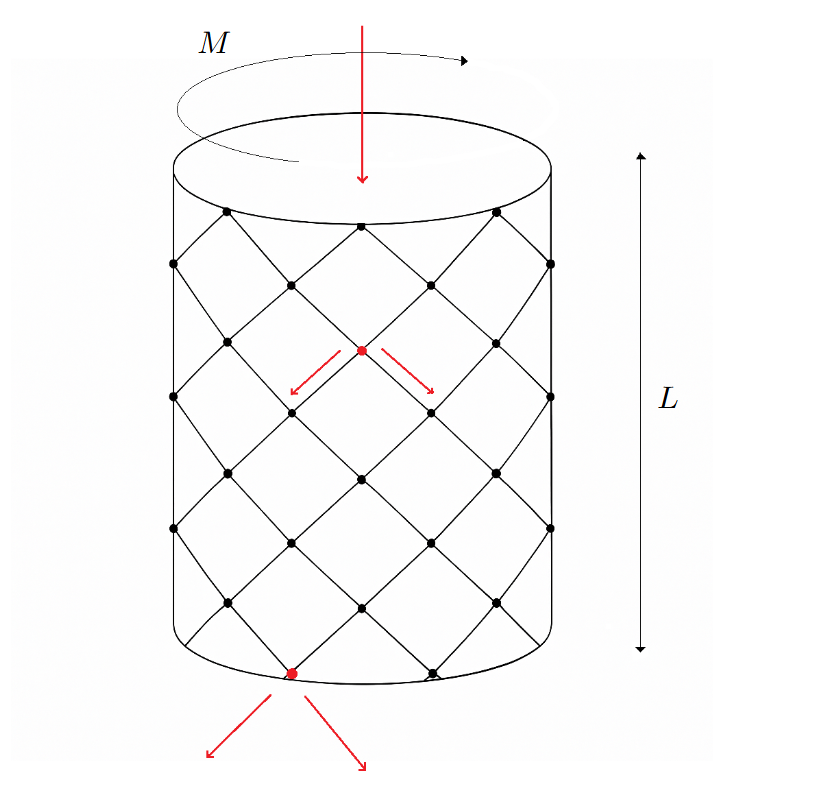}
    \caption{An $L\times M$ directed sandpile. Grains are added in the top layer. Unstable sites topple and grains move to the downward diagonal neighbours. Grains flow out from the bottom layer.}
    \label{cylinder}
\end{figure}

We consider the directed sandpile model \cite{directed} on a diagonal square lattice. The vertical height is $L$, and there are periodic boundary conditions in the horizontal direction. Each horizontal cross section has $M$ sites (Fig. \ref{cylinder}). At each lattice site $\vec{r}$ there is an associated number of sand grains $z_{\vec r} \geq 0$. If $z_{\vec r} \geq 2$, the site $\v r$ is said to be unstable. \\

At each time step, one sand grain is added at random to one of the sites $\vec r_0$ in the top layer, so that $z_{\vec r_0} \rightarrow z_{\vec r_0} +1$. If this makes the site unstable, the site is `toppled', and 2 sand grains are removed from $\vec r_0$, and 1 grain is transferred to each of the two immediate downward diagonal neighbours. In this manner, all unstable sites are successively toppled, until the sandpile reaches a stable configuration. Note that the toppling of a site in the bottom row results in 2 grains flowing out of the sandpile. The toppling moves are Abelian. A series of topplings of sites one below another, in the same timestep, is called an avalanche. The tractability of this model arises from the fact that each site topples at most once in an avalanche, and the avalanche progresses in a directed manner to the bottom of the sandpile. \\

Our analysis involves next timestep prediction of `large' avalanches in directed sandpiles. There are different ways to quantify the `size' of an avalanche. We denote the total number of toppled sites in an avalanche as $S_t$ and the vertical height of the avalanche as $h_t$. The number of topplings in the bottom row is called as the drop number, $f_t$. The net outflow of particles in a timestep $t$ is $2f_t$. We are given a time series of past data about sandpile events, upto time $T$. This could include observables (such as excess density $\ep_t$) or sizes of events.  Given this data we want to predict an extreme event at time $T+1$. We make binary predictions, wherein we turn on an `alarm' if we expect an extreme event to happen with appreciable probability, else we turn it off. \\

Notation : $\left(E_1, E_2, ..,E_T \right)$ is a binary time series where $E_i = 1$ if there is a large event at time $i$, and $E_i = 0$ otherwise. Similarly, $\left(X_1, X_2, ..,X_T \right)$ is the binary series of predictions made. At some time $t$, we set $X_{t+1} = 0$ (alarm off) or $X_{t+1} = 1$ (alarm on). Four possible outcomes arise with regard to our prediction : it is a true positive (tp), a true negative (tn), a false positive (fp) or a false negative (fn). The fractions for each of these are given by :
\begin{align}
    \text{tp} = T^{-1}\sum_{t=1}^{T}\delta_{E_t, 1}\delta_{X_t, 1} \, , \, \, \text{tn} = T^{-1}\sum_{t=1}^{T}\delta_{E_t, 0}\delta_{X_t, 0} \, , \, \,  \text{fp} = T^{-1}\sum_{t=1}^{T}\delta_{E_t, 0}\delta_{X_t, 1} \, , \, \, \text{fn} = T^{-1}\sum_{t=1}^{T}\delta_{E_t, 1}\delta_{X_t, 0} \, .
\end{align}

The prediction variables $\{ X_t\}$ are determined by the prediction strategy. Many prediction strategies are dependent on an intrinsic parameter $\sigma$ (a threshold) which can be chosen to optimize the quality of prediction. A popular class of curves that are plotted to judge the quality of predictions about a system are called ROC curves \cite{roc}. In such curves, the true positive rate $y = \frac{\text{fp}}{\text{tn +fp}}$ is plotted against the false negative rate $x = \frac{\text{fn}}{\text{tp + fn}}$, for a wide range of the parameter $\sigma$. The resulting curve is expected to be concave and strictly decreasing. \\

We prefer to use an equivalent, but more direct characterization, where on vertical axis we plot the fraction of positive events that are correctly predicted $c(\sigma) := \frac{\text{tp}}{\text{tp+fn}}$. This is plotted against the average fraction of time for which the alarm is on $\nu(\sigma) := \text{tp + fp}$. We will refer to this as an SROT (Success Rate versus On-Time)  curve. The goal is to get the best predictability for the least alarm on time. For an entirely unpredictable (uncorrelated) system, the SROT curve is a straight line $c = \nu$. For a partially predictable system, the SROT curve is expected to be convex and strictly increasing for $\nu \in [0,1]$.

\section{Strategy using waiting times}\label{waitingstrategy}
We assume that the signal (time series) containing information about extreme events is a stationary stochastic process for adiabatic driving rates (i.e. we wait for the configuration to stabilize before adding the next sand grain at the top).  In the directed sandpile, there is an anti-correlation between extreme events. For instance, a large avalanche at $t = 0$ drains out a lot of particles from the sandpile. So, another avalanche propagating through the same region is unlikely to be as big. Our strategy simply makes use of this low probability window to switch off the alarm. The strategy makes no use of any specific observables, but only the general statistical behaviour of the time series. We assume this statistics, in particular the distribution of waiting time intervals, is known. We define the following kinds of events as large: for toppling numbers, $S \geq S^* =  AL^{3/2}$, for heights $h \geq h^* = zL $, and for drop numbers $f \geq f^* = AL^{1/2}$. Here, $A,z$ are $\mathcal{O}(1)$ parameters. We call these events as `large' events because such avalanche sizes are much larger than the average values, which scale as $\overline{S} \sim L, \overline{h} \sim L^{1/2}$ and $\overline{f} \sim L^{0}$ for the directed sandpile. \\

Our strategy is as follows. Fix a time $\tau =\tau' L^{1/2}$ which is called the off time interval. Suppose there is a large event at time $t_0 = 0$. We switch off the alarm for the subsequent $\tau$ timesteps, and switch it back on at time $\tau + 1$. The alarm now remains on until another large event occurs.\\

We will consider the directed sandpile on an $L\times mL^{1/2}$ lattice. For large enough $m$, the local properties of the model are expected to be independent of $m$. This is because laterally distant sites in the model do not interact. The only effect of changing $m$ is that the effective timescale for local events is proportional to $m$.\\

We vary the off time parameter $\tau$, and use simulation data to obtain $c(\tau)$ and $\nu(\tau)$. We then plot SROT curves. 

\subsection{Monitoring large events globally}\label{globalevent}
Consider the $m \rightarrow \infty$ limit for our $L \times mL^{1/2}$ lattice. The time series available includes sizes of all avalanche occurring throughout the lattice. Henceforth, we call this the `global waiting time strategy', and call the monitored events as `global events'. Numerically, it is adequate to take $M = L$, because avalanches have widths that scale as $L^{1/2}$. We perform Monte Carlo simulations, and generate a time series (length $10^8$) of events for various $L$.

\subsubsection{Waiting time distributions for global events}
\begin{figure}
     \centering
     \begin{subfigure}{0.45\textwidth}
         \centering
         \includegraphics[width=\linewidth]{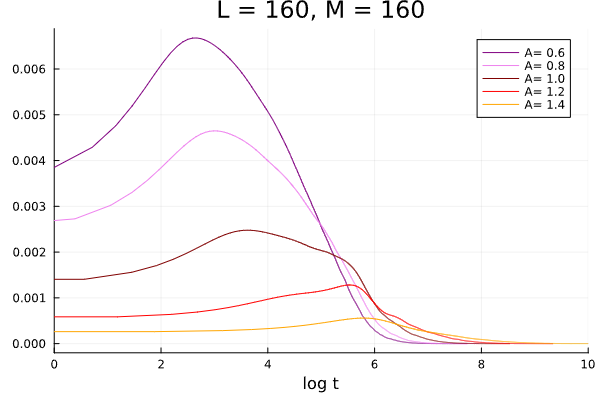}
         \caption{}
         \label{glob size wt 160}
     \end{subfigure}
     \begin{subfigure}{0.45\textwidth}
         \centering
         \includegraphics[width=\linewidth]{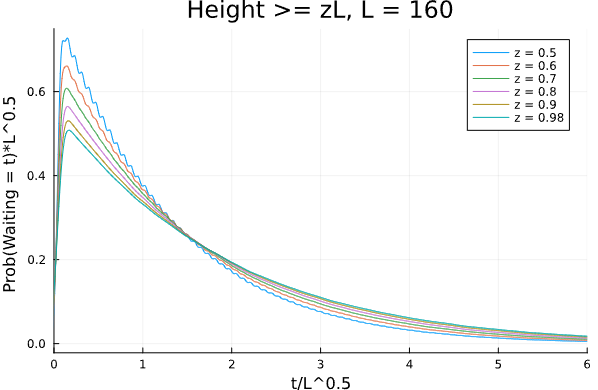}
         \caption{}
         \label{glob ht wt 160}
     \end{subfigure}
     \caption{Global event waiting time distributions for different $A$ for (a) Toppling number variables. Note that the $x$ axis is $\ln t$, this has been used because the distributions are extremely heavy tailed. (b) Height variables plotted for different thresholds $z$.}
\end{figure}

We plot the normalised waiting time distributions for `global events' of the form $S \geq AL^{3/2}$ and $h \geq zL$. Denote these distributions as $w_g^{S}(t)$ and $w_g^{h}(t)$. For a $160 \times 160 $ lattice and the above-mentioned time-series, the plots are in \ref{glob size wt 160} and \ref{glob ht wt 160} respectively.  \\

For large $A$ values the distribution is almost flat for a large range of $t$, because localized events far from the area of interest act as an effective noise for the waiting time distribution. The typical width of a big avalanche is $\m O\left(L^{1/2}\right) $, so a `local' region on the lattice around a point of interest can be defined as an $L\times L^{1/2}$ window. On average, the frequency of events outside such a local window is $ML^{-1/2}$ times that of events in the window. So, almost all the events in the `global' time-series are uncorrelated in the large $M$ limit, and the global waiting time distribution is simply an exponential \cite{garber, bofetta, flarepile}. This property is also seen in Fig. \ref{glob ht wt 160}.  
This global strategy cannot be used for prediction, because the flat nature of the global waiting time distribution does not give us a good threshold to switch off the alarm. \\

\begin{figure}
     \centering
     \begin{subfigure}{0.45\textwidth}
         \centering
         \includegraphics[width=\linewidth]{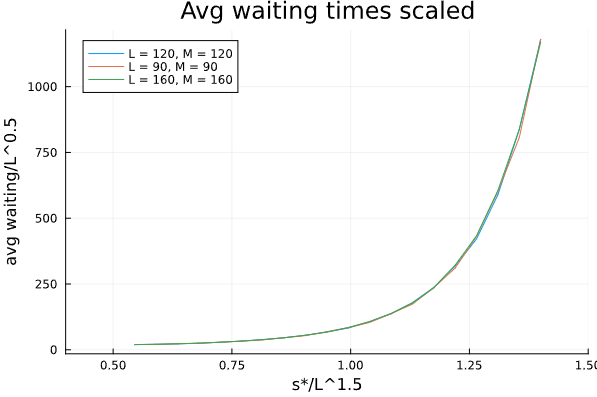}
         \caption{}
         \label{glob avg wt size}
     \end{subfigure}
     \begin{subfigure}{0.45\textwidth}
         \centering
         \includegraphics[width=\linewidth]{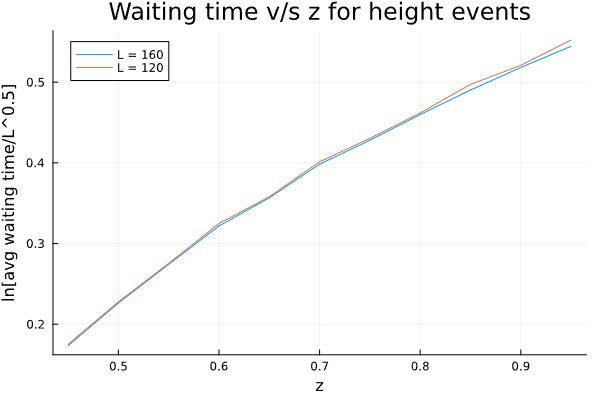}
         \caption{}
         \label{glob avg wt ht}
     \end{subfigure}
     \caption{Scaled average waiting times: (a) For toppling number events (b) For height events. Note that $\ln \left( L^{-1/2} \,\overline{t}_g^{h}\right)$ is plotted on the $y$ axis. }
\end{figure}

We also plot the average waiting time for toppling number events, $\overline{t_g^S}$ (as a function of $A$), and height events,  $\overline{t_g^h}$ (as a function of $z$). The plots are in Fig. \ref{glob avg wt size}, \ref{glob avg wt ht}, with $L = 90,120, 160$. As one would expect, we get a scaling collapse when the average waiting times are scaled by $L^{-1/2}$.

\subsubsection{SROT plots}
\begin{figure}
     \centering
     \begin{subfigure}{0.45\textwidth}
         \centering
         \includegraphics[width=\linewidth]{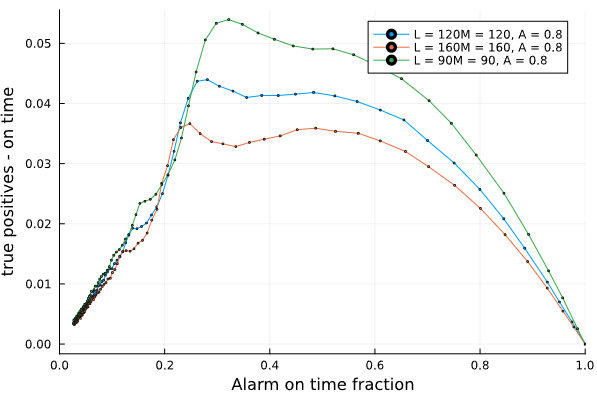}
         \caption{$A = 0.8$}
         \label{globwtroc0.8}
     \end{subfigure}
     \begin{subfigure}{0.45\textwidth}
         \centering
         \includegraphics[width=\linewidth]{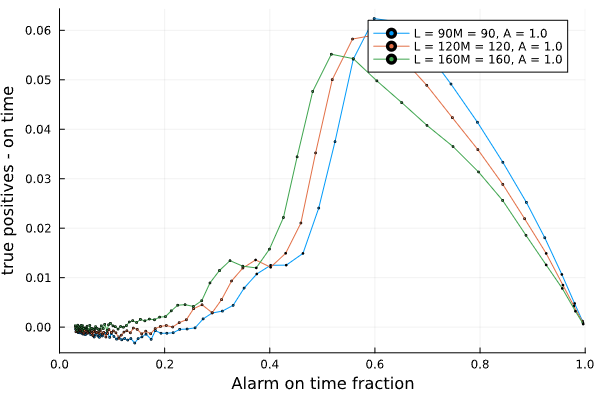}
         \caption{$A = 1$}
         \label{globwtroc1}
     \end{subfigure}
     \caption{Global waiting time strategy SROT plots for $M = L$.}
\end{figure}

It is clear that the predictability for global events is not going to be very good. So, instead of making the usual $c$ vs $\nu$ plot, we plot $c-\nu$ vs $\nu$ using the Monte Carlo data. We sample a wide range of waiting times $\tau$, and lattice sizes $L = 90,120,160$. The curves are shown in Fig. \ref{globwtroc0.8}, \ref{globwtroc1}. As expected, the prediction is not very effective; the gain in predictability is of $\m O\left(10^{-2}\right)$ for the considered lattice sizes.

\subsection{Local waiting time strategies}\label{localevent}

Consider the case where we have a local `detector', and only take into account events that occur within a finite distance of it. Given the noise in data with global events, it is clear that this will lead to  better predictions. We use a lattice of size $L \times mL^{1/2}$ with large enough $m$, and periodic boundary conditions. Within this, we consider a smaller vertical window of height $L$ and width $L^{1/2}$. We only take into account the outflow from the bottom row in this window. For simplicity, in this section we discuss the case when the prediction is based on drop number events $f \geq AL^{1/2}$. Note that only the outflow within the window is detected.

\subsubsection{Waiting time distributions for drop number variables}
We first analyse the waiting time distribution for such events. 
\begin{figure}
     \centering
     \begin{subfigure}{0.45\textwidth}
         \centering
         \includegraphics[width=\linewidth]{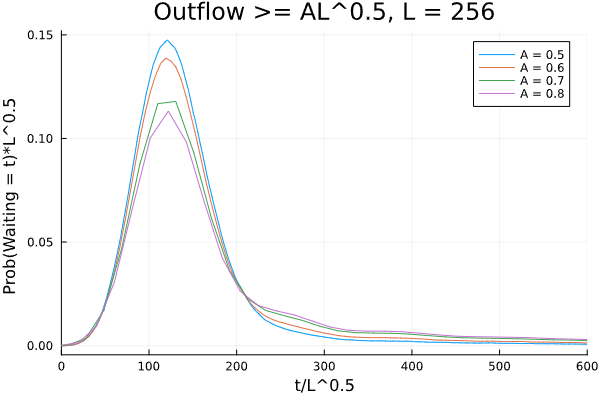}
         \caption{}
         \label{local L 256 var A}
     \end{subfigure}
     \begin{subfigure}{0.45\textwidth}
         \centering
         \includegraphics[width=\linewidth]{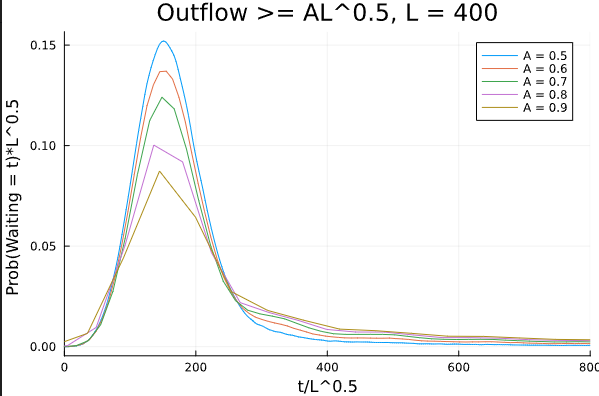}
         \caption{}
         \label{local L 400 var A}
     \end{subfigure}
     \caption{Waiting time distribution for local event $f \geq AL^{1/2}$ for varying $A$ and fixed $L=256, \, 400.$}
\end{figure}
In Fig. \ref{local L 256 var A} and \ref{local L 400 var A}, the $y$ axis shows the normalised waiting time distribution $w(t ; A,L)$ for the interval $t$ between successive large events. This is scaled by a factor of $L^{1/2}$, and is plotted against the scaled waiting time $tL^{-1/2}$. Notice that the graphs all begin from $(0,0)$ and peak at a time $t_m$. In each graph, the peak (mode), $t_m$ is independent of $A$. The height of this peak falls with $A$, because the distributions become increasingly long tailed for large $A$. 

\subsubsection{Average time spacing between events}
\begin{figure}
    \centering
    \includegraphics[width=0.5\linewidth]{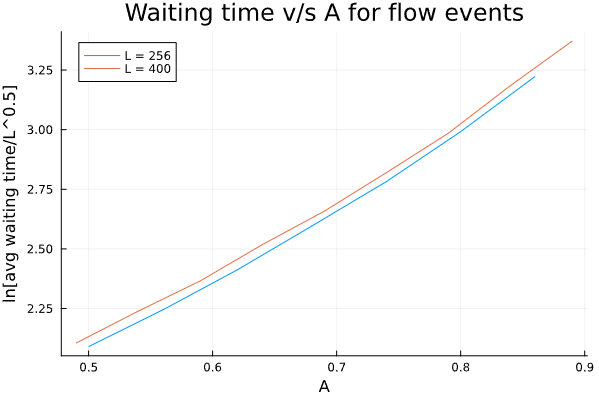}
    \caption{Average waiting time vs $A$, for $L = 256, \, 400.$}
    \label{local t_av vs A}
\end{figure}

In Fig. \ref{local t_av vs A} we plot the log of scaled average waiting time, $\ln \left(\overline{t}L^{-1/2} \right)$ against $A$, using the Monte Carlo data for $L=256,400$. There is a simple argument to get the dependence of $\overline{t}$ on $A$.\\

From the definition of the average time spacing between such `large' events, we have $\overline{t}^{-1} = \text{Prob}\,(f \geq AL^{1/2})$. Avalanches in the lattice will result in outflow throughout the bottom layer $x \in {1,..,mL^{1/2}}$ , but we are interested in the outflow in a smaller interval $1 \leq x \leq L^{1/2}$. Let the total outflow in the bottom layer be $f_g$, and that within our window be $f$. Then, in the large $L$ limit, the probability density for the scaled drop  number $v :=fL^{-1/2}$ is labelled as $h(v)$. With this, 
\begin{align}\label{tav}
    \overline{t}(A)^{-1} = L^{-1/2}\int_A^{1}dv \, h(v) \, .
\end{align}

It is a well known result for the 2-neighbour directed sandpile that the avalanche cluster has no holes. So, the two boundaries of the avalanche constitute an annihilating random walk. Suppose a particle is added at the top of the lattice at a co-ordinate $r_0$. Consider the annihilating random walk on the lattice corresponding to the ensuing avalanche. Let $x_1 < x_2$ be the co-ordinates of the two annihilating particles, and define $x = x_2-x_1 , \, r = (x_1+x_2)/2$. In these co-ordinates, the conditional density $\rho(x,r;y\,|\, 1,r_0;0)$ evolves as:
\begin{align}\label{anhw_full}
    \partial_y\rho (x,r ;y) = \frac{1}{2}\left[ \partial_r^2 + 4\partial_x^2 \right] \rho (x,r;y) \, .
\end{align}
Further, if we integrate out the centre of mass co-ordinate $r$ in (\ref{anhw_full}), we get the probability density $p(x;y|1,0)$ for a Brownian excursion \cite{redner}. The quantity $x(L)$ is the number of toppled sites in the bottom layer, so $x(L) = f_g$.  A well known result for Brownian excursions is that $p(x,y=L|1,0) = L^{-1/2}g_{b}\left(xL^{-1/2}\right)$, where $g_b(z) = \left( 4\pi\right)^{-1/2}z e^{-z^2/4}$. Define the intervals on $\mathbb{R}_{mL^{1/2}}$ : $I_0 = (0,L^{1/2}]$ and $I_{x,r} = \left[ r-\frac{x}{2} , r + \frac{x}{2} \right]$. Then we have:
\begin{align}
    h(u) = \int_{0}^{mL^{1/2}} \frac{dr_0}{mL^{1/2}} \int_0^{mL^{1/2}} dr \, dx \, \rho(x,r;L|1,r_0,0) \times \delta \left(uL^{1/2} - |I_{r,x} \cap L^{1/2}I_0| \right) \, .
\end{align}
The first factor of $1/mL^{1/2}$ is because $r_0$ is uniformly distributed. Using translation invariance, $\rho(x,r;L|1,r_0,0) = \rho(x,r-r_0,L|1,0,0)$. So, we get (after also rescaling all quantities by $L^{1/2}$, and defining the scaled interval $\ti I_{r',x'} = \left[ r' - \frac{x'}{2} , r' + \frac{x'}{2} \right]$ on $\mathbb{R}_m$):
\begin{align}
    h(u) = \frac{1}{m} \int_0^m  dx' \, g_b(x') \int_0^m dr' \, \delta \left(u - |\ti I_{r',x'} \cap I_0| \right) \, .
\end{align}
These integrals are easily evaluated. We get:
\begin{align}
    h(u) = \frac{e^{-u^2/4}}{2\sqrt{\pi}m}\left[ 4 + u(1-u) \right] + \frac{\delta(u-1)}{2 m} \,\text{erfc}\left( \frac{1}{2}\right) .
\end{align}
Finally, using (\ref{tav}),
\begin{align}\label{tavg}
    \overline{t}(A) = mL^{1/2}\left[ \text{erfc}\left(\frac{A}{2} \right)-\frac{1}{2}\text{erfc}\left(\frac{1}{2}\right)+\frac{A-1}{\sqrt{\pi}}e^{-A^2/4}\right]^{-1} \, .
\end{align}

\subsubsection{SROT curves for outflow variables}
\begin{figure}
     \centering
     \begin{subfigure}{0.45\textwidth}
         \centering
         \includegraphics[width=\linewidth]{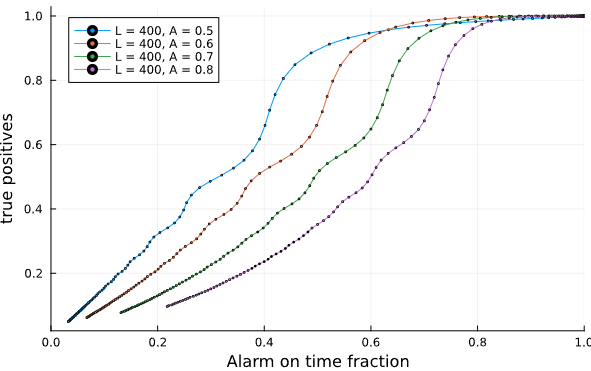}
         \caption{}
         \label{local wt ROC 400}
     \end{subfigure}
     \begin{subfigure}{0.45\textwidth}
         \centering
         \includegraphics[width=\linewidth]{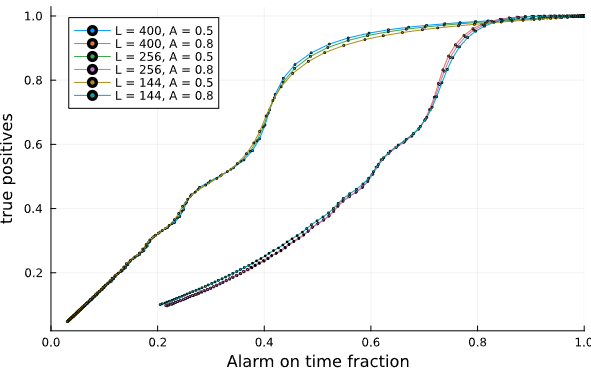}
         \caption{}
         \label{local wt ROC collapse}
     \end{subfigure}
     \caption{SROT plots. (a) Plots for $0.5 \leq A \leq 0.8$ (b) Curves are identical for $L = 144, \, 256, \, 400$ with $A = 0.5, \, 0.8$.}
\end{figure}

Here we discuss MC results and plot SROT curves. The plots are generated using the time series for $L = 144,256,400$ and a multiple values of $A$ (Fig. \ref{local wt ROC 400}, \ref{local wt ROC collapse}).
We make the following observations. Firstly, for a given $A$, there is a scaling collapse in the SROT plots for a very wide range of $L$. At large lengths, there are only small deviations from the true thermodynamic limit; therefore there is finite predictability in the thermodynamic limit. \\

Next, there is an approximate `on time' fraction $\nu^*$ such that for $\nu > \nu^*$, $c$ is close to 1. The value of $\nu^*$ increases with $A$. Plots for larger $A$ are similar to those for smaller $A$, except that they are `pushed' to the right. To understand this we examine the waiting time distribution (Fig. \ref{local L 400 var A}). Fix $L$ and $A$. The function $w(t,A,L)$ is peaked about $t_m = t_m'L^{1/2}$. If the alarm is switched off for a time $\tau < t_m$, then the average number of large events occurring in the of interval is pretty small, and nearly all avalanches are predicted. This corresponds to $\nu > \nu_m^*$ in Fig. \ref{local wt ROC collapse}. Once we cross the peak, i.e. $\tau>t_m$ and $\nu < \nu^*$, there is a surge in the number of missed avalanches and the predicted fraction suddenly drops. On increasing $A$, the height of the peak, $w(t_m)$ decreases and $w$ becomes more and more heavy tailed. Suppose we keep the alarm off for a fraction $\tau < t_m$ of the time. Then, the alarm is switched on until the next large event. Due to the heavy tailed nature of $w$, the next large event takes a much longer time to occur on average. So the on time fraction is much larger, and the ROC curve is pushed to the right for larger $A$. This means that very extreme events ($A \sim 1$) are in this sense `less' predictable, which is different from other results \cite{shapoval, garber}, but not contradictory. The result is specific to the current strategy. In fact, for subsequent strategies, we will see that predictability improves with rarer and larger events. Moreover, it is also true that for $\nu > \nu^*$, the $c$ is larger for larger $A$, though all $c$ values are close to 1.\\

A possible extension of this strategy involves using the conditional waiting time distributions given the size of the previous large event. This may improve the prediction efficiency.

\subsubsection{What are the SROT curves ?}
Under the assumption that the time intervals between successive large events are i.i.d variables, and that their distribution $w(t)$ is known, we can calculate the SROT curves. Fix the alarm-off threshold $\tau$. Suppose that an event happens at $t_0 = 0$. There can be multiple `missed' events in the interval $(0,\tau)$. Suppose there are $n$ of them, at times $t_1\leq t_2 \leq ..\leq t_n$. Let $t_{n+1} > \tau$ be the next `detected' event. Since the time evolution is statistically stationary, it suffices to consider the statistics between two detected events. The $c$ and $\nu$ parameters (true positive fraction and alarm on time fraction) for the SROT curves can be written as :
\begin{align}
    c(\tau) &= \sum_{n=0}^{\infty} \frac{1}{n+1} \int_{\tau}^{\infty} dt_{n+1} \times \text{Prob}(n \text{ events in } (0, \tau) \, | \, n+1 ^\text{th} \text{ event at }t_{n+1} ) \nonumber \\
    \nu(\tau) &= 1 - \tau\sum_{n=0}^{\infty} \int_{\tau}^{\infty} \frac{dt_{n+1}}{t_{n+1}} \times  \text{Prob}(n \text{ events in } (0, \tau) \, | \, n+1 ^\text{th} \text{ event at }t_{n+1} ) \, .
\end{align}
Let $\ti{c}(r) := \int_{0}^{\infty } d\tau e^{-r\tau}c(\tau)$ and $\ti{\nu}(r) := \int_{0}^{\infty } d\tau e^{-r\tau}\nu(\tau)$ be the Laplace transforms of $c(\tau), \nu(\tau)$. From the i.i.d property we have :
\begin{align}
    \text{Prob}(n \text{ events in } (0, \tau) \, | \, n+1 ^\text{th} \text{ event at }t_{n+1} ) = \int_{0}^{\tau} dt_1 \int_{t_1}^{\tau} dt_2 \, ...\, \int_{t_{n-1}}^{\tau} dt_{n} \, w(t_{n+1}-t_n)\prod_{j = 0}^{n-1} w(t_{j+1} -t_{j}) \, .
\end{align}
Let $q(t_n) := \text{prob} (n-1 \text{ events in } (0,t_n) )$ where the $n^\text{th}$ event happens at $t_n$. Then,
\begin{align}
    q(t) &= \int_{0}^{t_n} dt_1  ..\, \int_{t_{n-2}}^{t_n} dt_{n-1} \, \prod_{j = 0}^{n-1} w(t_{j+1} -t_{j}) 
    \Rightarrow \ti q(s) = \left[ \ti w (s)\right]^n \, . 
\end{align}
Above $\ti q(s)$ is the Laplace transform of $q(t)$. Let us first evaluate $c$.
\begin{align}\label{c(r)}
    \ti c (r) &= \int_0^\infty d\tau\,  e^{-r\tau } \sum_{n=0}^{\infty}\frac{1}{n+1} \int_{\tau }^{\infty} dt_{n+1} \int_{0}^{\tau } dt_n \, q(t_n) \, w(t_{n+1} - t_n) \nonumber \\
    &= \frac{1-\ti w(r)}{r} \times \int^* \frac{ds}{2\pi i} \frac{\ln \left[1-\ti w(s) \right]}{\ti w (s)(r-s)} \, .
\end{align} 
In the second line we have used a Mellin integral for the inverse Laplace transform. Next, we evaluate $\ti \nu(r)$. 
\begin{align}\label{nu(r)}
    \ti \nu  (r) &= \int_0^{\infty}d\tau  \, \tau e^{-r\tau}\sum_{n=0}^{\infty}\int_\tau^{\infty} \frac{dt_{n+1}} {t_{n+1}} \int_0^{\tau} dt_n \, w(t_{n+1}-t_n) \, q(t_n) \nonumber \\ 
    &= \frac{1}{r}\int^* \frac{ds}{2\pi i} \frac{1}{1-\ti w(s)} \, \left[ \frac{\ti w (r)}{s-r}-\int_{r}^{\infty} \frac{dk \, \ti w(k)}{r(k-s)} + \int_0^\infty\frac{dk \, \ti w(k)\, (k+2r-s)}{r(k+r-s)^2}\right] \, .
\end{align}

\subsubsection{Special limits}

Next we check the limiting behaviour of the SROT plots, i.e. the top right and bottom left regions of Fig. \ref{local wt ROC 400}. These correspond to $\tau \ll t_m$ and $\tau \gg t_m$ respectively, where $t_m = t_m'L^{1/2}$ is the mode of $w(t)$. Let $\tau = \tau'L^{1/2}$. \\
In the first case, the alarm is switched off only for a short interval corresponding to the $t' \ll t_m'$ region in Fig. \ref{local L 256 var A}. Let $w\left(t'L^{1/2}\right)_{t' \ll t_m'} \sim w_0(t')$. For instance, it may be a power law. In the short off interval, the probability of $n$ missed events falls very quickly with $n$, so we can keep only the first few small $n$ terms in the calculation of $c$ and $\nu$.
\begin{align}
    c(\tau) \simeq \int_{\tau}^{\infty} \, dt_1 w(t_1) +\frac{1}{2} \int_{\tau}^{\infty} dt_2 \int_{0}^{\tau} dt_1 \, w(t_2-t_1)w(t_1)  &\simeq 1-\frac{1}{2}\tau w_o(\tau) . \nonumber \\
    1-\nu(\tau) \simeq \tau \int_{\tau}^{\infty} dt_1  \, t_{1}^{-1} w(t_1) + \text{H.O.} &\simeq \ \frac{\tau}{t_m} = \frac{\tau'}{t_m'}\, .
\end{align}
With $\overline{c} = 1-c , \, \overline{\nu} = 1-\nu$, the small $\tau $ behaviour of the plot is :
\begin{align}\label{asym1}
    \overline{c} \simeq \frac{1}{2} \overline\nu \, t_m' \, L^{-1/2}\, w_0\left(\overline\nu \, t_m'\right) \, .
\end{align}
For instance, if $w_0(t') \sim t'^{\gamma}$, $\overline{c} \sim \overline{\nu}^{\gamma +1}$. Note that in the thermodynamic limit, $\overline{c} \rightarrow 0$. \\
In the opposite limit, the alarm is switched off for a very large time $\tau \gg t_m$. The distribution for the number of avalanches in the interval $(0, \tau)$ is sharply peaked about $n^* = \tau/t_m$. (This can be shown by taking a Gaussian approximation of $w(t)$ about $t_m$, in which case only the events around the peak $t_m$ dominate the probability measure). We have :
\begin{align}\label{asym2}
    c &\simeq \frac{t_m}{\tau} \, , \, \nu \simeq 1-\tau \int_0^{\infty} \frac{ dt \, w(t)}{t + \tau} \simeq \frac{\overline{t}}{\tau} \, . \nonumber \\
    c &\sim \frac{t_m}{\overline{t}} \nu \, .
\end{align}
Above, $\overline{t} \equiv \overline{t}(A)$ is the average spacing time given in (\ref{tavg}). For increasing $A$, we know that $\overline{t}(A)$ increases, whereas $t_m$ is independent of $A$. One can see in the plots (Fig. \ref{local wt ROC collapse}) that for larger $A$, the slope of the curve near $(0,0)$ is diminished.

\subsubsection{Other local variables : toppling numbers and heights.}
\begin{figure}
    \centering
    \includegraphics[width=0.5\linewidth]{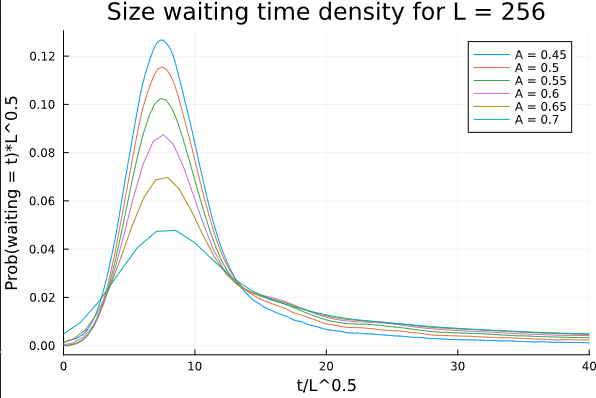}
    \caption{Local waiting time distribution for $S \geq AL^{3/2}$.}
    \label{local wt size 256}
\end{figure}

The previous sections have dealt with drop number variables. But the equations (\ref{c(r)}), (\ref{nu(r)}), (\ref{asym1}) and (\ref{asym2}) were derived for an arbitrary waiting time distribution $w(t)$. So, they hold good also for toppling number and height events. Of course, we must use the appropriate waiting time distributions $w^{S}(t)$ and $w^h(t)$ in the two cases. From the time series data, we plot the local waiting time distributions for toppling number events $S \geq AL^{3/2}$, with $L=256$ and a range of $A$ (Fig. \ref{local wt size 256}). Similar to the outflow curves, the peak in these curves is also independent of $A$. 

\section{Strategies using the average excess density in the pile}\label{xsdens}

The sandpile model involves a Markov update rule for the full configuration. The time series $\{ S_t\} , \{h_t\} \text{ and } \{f_t\}$ constitute Hidden Markov processes \cite{hiddenmarkov}. To enhance the predictability of the observables, one has to incorporate some more information about the sandpile configuration. We will use the excess density $\epsilon \in (-1,1)$, which is defined by the total mass on the lattice being $N(1+\epsilon)/2$. We are given a past-time series $\ep_1, \ep_2,..,\ep_t$. We want to predict events of the form $S \geq AL^{\gamma}$, $h \geq zL^{\bt}$ or $f \geq AL^{\delta}$ where $z, A \sim \m O(1)$, and $\gamma, \bt,\delta $ are appropriate scaling exponents. The strategy fixes the prediction variable as :
\begin{align}
    X_{t+1} = \begin{cases}
        1 , & \ep_t \geq \ep^* \\
        0, & \ep_t < \ep^* \, .
    \end{cases}
\end{align}
Two trivial cases can be immediately understood. In the case $\ep^* \rightarrow -1$, the alarm is always on. Consequently, all avalanches are predicted and $c = \nu = 1$. Conversely, for $\ep^* \rightarrow +1$ the threshold is too high and the alarm is almost always off. So $c = \nu = 0$. We can vary the parameter $\ep^*$ in $[-1,1]$ to get SROT curves $c(\ep^*)$ vs $\nu(\ep^*)$.

\subsection{Height variables}\label{xsheight}
We discuss the behaviour of predictions for events of the form $h \geq zh^*$.
\subsubsection{Scaling theory} \label{hscal}
We have the unscaled variables $h^*, \ep $. Let us define scaling exponents so that scaled variables are $z := h^*L^{-\al}\, , \xi := \ep L^{\bt}$. Further, set $M = mL^{\mu}$. In this strategy, the prediction for time $t+1$ is blind to all data other than that at $t$. So, one must average over all possible histories of the system  and the measure that we use is same as the steady state measure. The $x$ and $y$ axis of our SROT curve are given by :
\begin{align}
    c &= \frac{\text{Prob} \left(h \geq zL^{\al} \, ; \, \ep \geq \xi L^{-\bt} \, \right)}{\text{Prob} \left( h \geq zL^{\al}\right)} \equiv \frac{1}{\text{Prob} \left( h \geq zL^{\al}\right)} \int_{\xi L^{-\bt}} \text{Prob}(d \ep') \times \text{Prob} \left(h \geq zL^{\al} \,| \, \ep' \right) \, . \\
    \nu &= \text{Prob} \left( \ep \geq \xi L^{-\bt} \, \right) \, .
\end{align}
All integrals are over the steady state measure $\text{Prob}(d \ep)$. In the steady state all states are equally probable. This is due to the fact that in the directed model, all states are recurrent \cite{recurrent}. So, the probability to have an excess mass $N\ep /2$ in the system is simply $2^{-2N}\binom{N}{\frac{N}{2} (1+\ep)}$. Here, $N := L^{1+\mu}$ is the total number of lattice sites. In the thermodynamic limit, the density is:
\begin{align}\label{density}
    \rho \left(\ep = \xi L^{-\bt}\right) \sim \frac{m^{-\frac{1}{2}}L^{-\frac{1+\mu}{2}}}{\sqrt{2\pi}}\exp \left( -\frac{m\xi^2}{2} L^{1+\mu-2\bt}\right) \, .    
\end{align}
The typical density fluctuation is thus of the order $L^{-(1+\mu)/2}$. So if we set $2 \bt = 1+ \mu$, we get a scaling form $\rho(\ep = \xi L^{-\bt}) = L^{-\bt} \, \varrho(\xi)$.\\

We now compute $\text{Prob} \left(h \geq zL^{\al} \,| \, \ep \right)$. In the directed model, the boundary of the set of toppled sites in every avalanche can be mapped to the path of a pair of annihilating random walkers in one dimension. Let $x(t) = x_2(t)-x_1(t)$ be the distance between the two walkers. Here we set $x_2 > x_1 \, , \, \forall t \in \{1,..,h-1\}$, and $x_2(h) = x_1(h)$. Thus the walkers annihilate at a vertical height $h$. \\

Given a uniform average excess density $\ep$, the jump probabilities are given by:
\begin{align}\label{jumprob}
    \left(\frac{1\pm \ep}{2}\right)^2 \text{ for a forward/backward jump (}x_{i+1}-x_i = \pm 1)\, ; \,  \frac{1-\ep^2}{2} \text{ for no jump } (x_{i+1} = x_i).
\end{align}
The probability density $p(x; t)$ satisfies:
\begin{align}\label{anhw}
    p(x;t+1) = \left( \frac{1+\ep}{2}\right)^2 p(x-1; t) +\left( \frac{1-\ep}{2}\right)^2 p(x+1; t) +\frac{1-\ep^2}{2} p(x; t) \,.
\end{align}
Also,
\begin{align}
    \text{Prob} \left(h \geq zL^{\al} \,| \, \ep \right) = \text{Prob} \left( x(h) > 0 \text{ in the annihilating RW}\right) \,.
\end{align}
The walker begins at $x(0) = 1$, remains in the region $x > 0$ and makes a first passage across $x = 0$ at time $h$. We must calculate the first passage distribution $Q_0(t|x)$ for the RW defined by (\ref{anhw}). Here, $Q_0(t|x)$ is the probability that the walker makes the first passage across $0$ at time $t$, given that it is initially at point $x$. First, note the convolution property :
\begin{align}
    Q_0(t|x) = \sum_{u=0}^{t}Q_0(t-u|x')Q_{x'}(u|x) = \sum_{u=0}^{t}Q_0(t-u|x')Q_0(u|x-x')\, , \text{where } 0 < x' <x.
\end{align}
The second equality is true because of translational invariance. Defining the generating function $q(z|x) := \sum_{t \geq 0}z^tQ_{0}(t|x)$, we directly get :
\begin{align}\label{fpr1}
    q(z|x) = q(z|1)^x \, .
\end{align}
Next, suppose that the particle starts off from point $1$. Using (\ref{anhw}):
\begin{align}
    Q_0(t|1) &= \frac{(1-\ep)^2}{4}\,\delta_{t,1} + \frac{(1+\ep)^2}{4} \, Q_0(t-1|2) + \frac{1-\ep^2}{2}Q_0(t-1|1) \,.
\end{align}
Writing this in terms of the generating functions $q(z|x)$ and using (\ref{fpr1}), we get a quadratic in $q(z|1)$. This is easily solved. 
\begin{align}
    q(z|1) = \frac{2z^{-1}}{(1+\ep)^2} \left[ 1- \frac{1-\ep^2}{2} z - \sqrt{1-(1-\ep^2)z} \, \right] \, .
\end{align}
Expanding the square root one can extract the coefficient of $z^t$ and obtain $Q_0(t|1)$, which is relevant to us because our walker starts at $x = 1$ at $t = 0$. For large $h$ one can take a continuum limit :
\begin{align}\label{con}
    \text{Prob} \left(h \geq zL^{\al} \,| \, \ep \right) = \sum_{t = zL^\al}^{\infty}Q_0(t|1) \sim \frac{1}{(1+\ep)^2} \left[ 4\ep \Theta(\ep) + \frac{1}{2 \sqrt{\pi}}\int_{zL^{\al}}^{\infty} dy \, y^{-3/2} \left(1-\ep^2 \right)^{y} \right] \, .
\end{align}
We can now use the scaling form $\ep = \xi L^{-\bt}$. Since $\bt > 0 $ we can ignore terms like $(1+\ep)^{-2}$ and so on. We get :
\begin{align}
    \text{Prob} \left(h \geq zL^{\al} \,| \, \ep \right) = L^{-\bt}\left[ 4  \xi \Theta(\xi) + \frac{L^{\bt - \al/2}}{2 \sqrt{\pi z}} \int_{1}^{\infty} dy \, y^{-3/2} \exp \left( -\xi^2 yzL^{\al - 2\bt}\right) \right] \, .
\end{align}
If we set $\al = 2\bt$ the $L$ dependence in the integral vanishes and we get a scaling form $\text{Prob} \left(h \geq zL^{\al} \,| \, \ep  = \xi L^{-\bt}\right) = L^{-\bt}g(\xi, z)$. Here, $g(\xi, z) := 4\xi\Theta(\xi) + (4 \pi z)^{-1/2} \int_1^{\infty}dy \, y^{-3/2} \exp\left( -\xi^2yz \right)$ .\\

\begin{figure}
    \centering
    \includegraphics[width=0.45\linewidth]{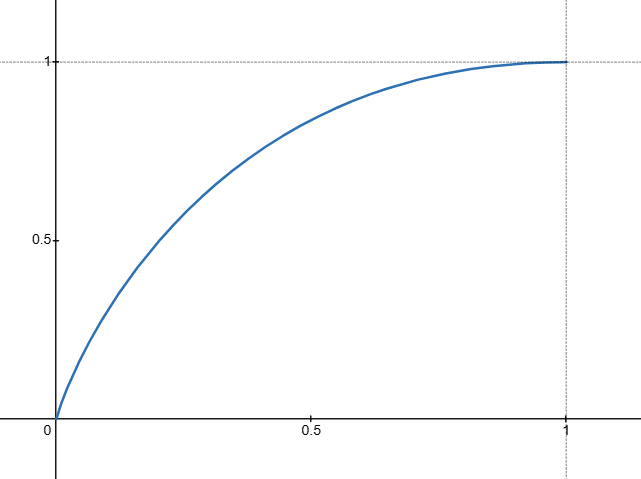}
    \caption{SROT curve for $m=2$, $z$ = 0.8}
    \label{par}
\end{figure}

We return to the calculation of the $c$ and $\nu$ variables.
\begin{align}\label{xpar}
    \nu(\xi) := \text{Prob} \left( \ep \geq \xi L^{-\bt}\right) \rightarrow (2 \pi )^{-1/2}m^{1/2} \int_{\xi}^{\infty} d \xi' e^{-m\xi'^2/2} = \frac{1}{2} \text{erfc} \left( \xi \sqrt{\frac{m}{2}}\right) \, .
\end{align}
Let us define $\phi(\xi, z) = \int_{\xi}^{\infty} d\xi' \, \varrho(\xi') \,  g(\xi',z)$. Then :
\begin{align}
    \text{Prob} \left(h \geq zL^{\al} \, ; \, \ep \geq \xi L^{-\bt} \, \right) &= L^{-2\bt} \phi(\xi, z) \, , \, \text{Prob} \left(h \geq zL^{\al} \right) = L^{-2 \bt} \phi (-\infty, z). \nonumber \\
    \Rightarrow c(\xi) &= \frac{\phi(\xi, z)}{\phi(-\infty,z)} \, .
\end{align}
One can perform the integrals in the expression for  $\phi(\xi, z)$ and write it as :
\begin{align}\label{ypar}
    c(\xi) = \frac{1}{2} \text{erfc} \left( \xi \sqrt{z+\frac{m}{2}} \right) + \frac{1}{2} \sqrt{\frac{z}{z+\frac{m}{2}}}\, e^{-m\xi^2/2} \left[ 1 + \text{erf} \left( \xi \sqrt{z}\right)\right] \, .
\end{align}
It is crucial that we have a continuous expression about $\xi = 0$, despite having different expressions for $g(\xi)$ for $\xi > 0$ and $\xi < 0$. This is because, in the final SROT curve, the $\xi = 0$ point corresponds to some nontrivial $x, \, 0 <x<1$, where the alarm is on for a finite fraction of time. We would want a continuous SROT curve about this point. With (\ref{xpar}) and (\ref{ypar}) a parametric curve is plotted in Fig. \ref{par}. On decreasing $z$, the curve moves closer to $y = x$. This means that rarer events are more easily predictable. \\

The relations between our scaling exponents are given by $\al = 2\bt = 1+\mu$. For $\mu = 1/2$, this implies that $\al =3/2>1$, and that the scaling theory of large events can be only used if events are larger than $\m O \left( L^{3/2} \right)$.(We get a scale invariant prediction curve for events of the form $h \geq zL^{3/2}$.) But the physical bound on avalanche heights is $h \leq L$, and the relations between the scaling exponents cannot be satisfied for a directed sandpile.  \\

The global excess density shows Gaussian fluctuations, and the typical value is of the order $\ep_0 \sim L^{-(1+\mu)/2} = L^{-3/4}$. It turns out that for large $L$, such values of excess density are too small to significantly change the distribution of local avalanches. Height events are well predicted for excess densities of order $L^{-1/2}$, but these large density values are very rarely realised in the set of recurrent states of the directed sandpile. \\

\begin{figure}
    \centering
    \includegraphics[width=0.45\linewidth]{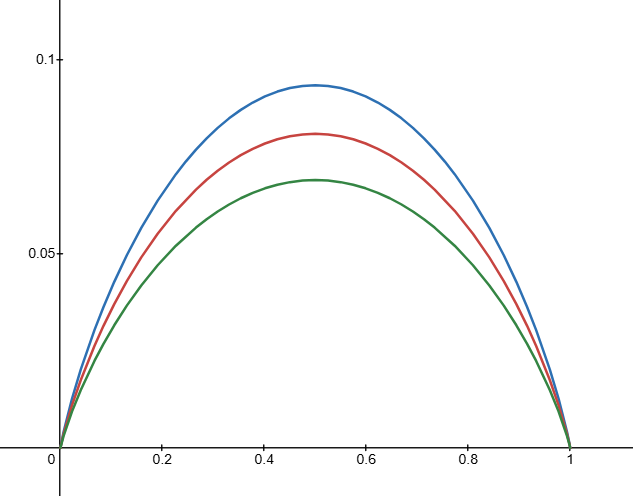}
    \caption{Plots of $c-\nu$ vs $\nu$ for $L = $ 144 (blue), 256 (red), 484 (green).}
    \label{h finite parametric}
\end{figure}

In a sandpile with large but finite $L$, height variables are weakly predictable, as the numerics show. The natural scaling for height predictions is of the form $h \geq h^* = zL$, where $z \in (0,1]$. With $\mu = 1/2$ (i.e. $M = mL^{1/2}$), one can repeat the calculation in the previous section to get the parametric equation for the SROT curves. With $\ep^*  = \xi^*L^{-3/4}$, and $\xi^* \in (-\infty, \infty)$ as the parameter, the equations are :
\begin{align} \label{finite1}
    \nu(\xi) &= \frac{1}{2} \text{erfc} \left( \xi \sqrt{\frac{m}{2}}\right) \nonumber \\
    c(\xi ) &=  \frac{1}{2} \text{erfc} \left( \xi \sqrt{zL^{-1/2}+\frac{m}{2}} \right) + \frac{1}{2} \sqrt{\frac{2zL^{-1/2}}{2zL^{-1/2}+m}}\, e^{-m\xi^2/2} \left[ 1 + \text{erf} \left( \xi \sqrt{zL^{-1/2}}\right)\right] \, .
\end{align} 
Expanding $c(\xi) - \nu(\xi)$ for large $L$, we get :
\begin{align}\label{finite 2}
    c(\xi)-\nu(\xi) \simeq \sqrt{\frac{z}{2m}}L^{-1/4} \, e^{-m\xi^2/2} + \mathcal{O}\left (L^{-1/2} \right) .
\end{align}
The $c-\nu$ vs $\nu$ curve defined by (\ref{finite1}, \ref{finite 2}) is plotted in Fig. \ref{h finite parametric}. It has a peak at $\xi = 0$, which scales as $L^{-1/4}$. Also, note that the quantity $c-\nu$ is depends on the combination $zL^{-1/2}$ in (\ref{finite 2}). This means that the $c-\nu$ curves should collapse when we vary $z$ as $L^{1/2}$, i.e. $h^* \sim L^{3/2}$. We see that in the limit $L \rightarrow \infty$, the SROT curve asymptotically reaches $c = \nu$. Thus the system is not predictable in the thermodynamic limit. \\

We have restricted ourselves to studying the directed sandpile with two diagonal downward neighbours. However, we expect these results to hold for more general directed sandpiles with multiple downward neighbours \cite{multipleneighbour}. This is because the cutoff on heights and toppling numbers of avalanches due to the directed-ness of the model holds in the latter case too.

\subsubsection{Numerical plots}\label{hnum}
For numerical simulations we first choose an $L \times 4L^{1/2}$ lattice and consider a smaller $L \times L^{1/2}$ window within it. For all plots we use a time series (length $10^8$). We monitor the excess density in the window, and local events of the form $h \geq AL$. 

\begin{figure}
     \centering
     \begin{subfigure}{0.45\textwidth}
         \centering
         \includegraphics[width=\linewidth]{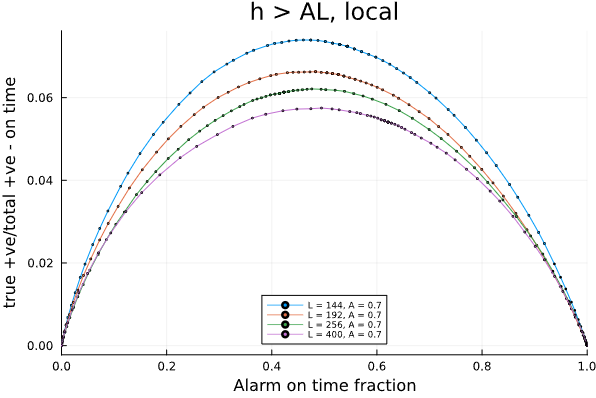}
         \caption{A = 0.7}
         \label{glob size wt}
     \end{subfigure}
     \begin{subfigure}{0.45\textwidth}
         \centering
         \includegraphics[width=\linewidth]{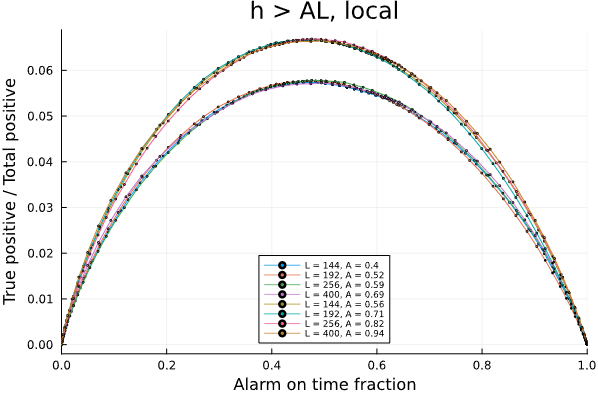}
         \caption{$A \sim L^{0.39}$}
         \label{glob ht wt 120}
     \end{subfigure}
     \caption{(a) Curves are distinct for different lengths $L = $ 144, 192, 256, 400 and fixed $A = 0.7$. (b) Data collapses for $A \sim L^{0.39}$, for two different sets of values of $A$, and $L = $ 144, 192, 256, 400. }
\end{figure}

The plots indicate that the effective scaling form for finite $L$ is $A \sim L^{0.39}$. This should not come as a surprise, because this means that $h^* \sim L^{1.39}$, which is similar to the exponent $\al = 1+\mu = 3/2$ obtained from the scaling theory. The height variables are not predictable in the thermodynamic limit, and the plot asymptotically tends to $y = x$.

We have also carried out some direct simulations of the height variables on an $L\times 2L^{1/2}$ lattice with periodic boundaries, without taking a window. In this case, the SROT curves collapse is when we scale $A \sim M^{0.69} \sim L^{0.35}$ so that $h^{*} \sim L^{1.35}$. This differs from the $3/2$ exponent above, because periodic boundaries with smaller $M$ can lead to more frequent `persistent avalanches, where all the sites in a given layer topple. The effect of such avalanches is an artifact of the special choice of aspect ratio of the lattice.

\subsection{Toppling number variables}\label{xssize}
The events we predict are of the form $S \geq AL^{3/2}$. Again, we plot $(c-\nu)$ vs $\nu$ for better clarity.

\subsubsection{Numerical SROT plots}

\begin{figure}
    \centering
    \includegraphics[width=0.5\linewidth]{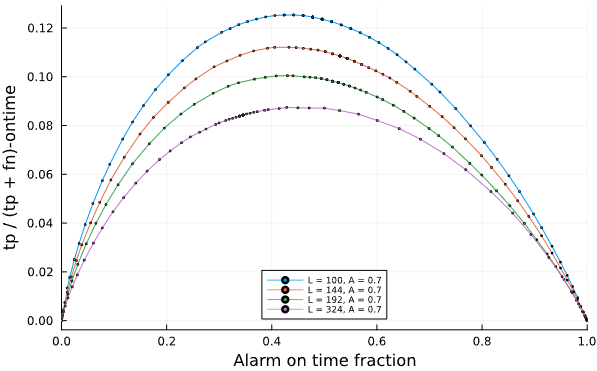}
    \caption{$c-\nu$ vs $\nu$ plots for size events. $A = 0.7$ is fixed, and $L = 100, 144, 192, 324$.  }     
    \label{size window all}
\end{figure}

We first fix $A$ and use the Monte Carlo data to get SROT type plots, for $L\times 4L^{1/2}$ lattice sizes with an $L \times 2L^{1/2}$ window. The size is monitored within this window. As usual, we plot the quantities $c-\nu$ against $\nu$. For a fixed $A = 0.7$, we get distinct plots for different lattice sizes (Fig. \ref{size window all}). \\

We saw in Sec. \ref{hscal}, \ref{hnum}, that for height events of the form $h \geq zL$, the $c-\nu$ vs $\nu$ plots collapse for $z \sim L^{1/2}$. For toppling number events of the form $S \geq AL^{3/2}$, however, we do not find a simple power law scaling of $A$ with $L$ for the curves to collapse. This is unlike the height case, where $z \sim L^{1/2}$ works. We will show in Sec. \ref{sizeroc} that the form of the SROT plots depend only on a quantity $L^{-1/4}g(A)$, where $g(A)$ is a non-trivial function. No simple power law scaling of $A$ with $L$ is possible. \\

We want to have an analytical understanding of the above curves, similar to that for the height case. For this, we first need to derive the distribution for avalanche sizes.

\subsubsection{Probability distribution for avalanche toppling numbers}

We consider an $L \times mL^{1/2}$ lattice with an excess mass density $\ep$. We would like to obtain the probability $\m P ^{\ep}(S,L)$, i.e. the probability that an avalanche has a toppling number $S$ with this density.\\

We can again use the fact that the boundaries of the avalanche constitute an annihilating random walk. In particular, we have a single walker in the discrete half space $x \in \mathbb{Z}_+$, with an absorbing boundary at $x = 0$. The probability density for this random walk satisfies the dynamical equation (\ref{anhw}). We have $S = \sum_{j=1}^L x_j$. For a lattice of height $L$, there are two possibilities for the fate of the walker at height $L$. The particle could have been absorbed at some height $t < L$, in which case $S = \sum_{j=1}^{t}x_j$; or it could be at some finite co-ordinate $x_L = b$. Given these cases we have :
\begin{align}\label{sizedisc}
    \m P^{\ep}(S,L) = \sum_{t=1}^{L} P^{\ep}(S |x_t=0 , \, x_0 = 1) \, \text{Prob}(h = t |\ep) + \sum_{b>0} P^{\ep}(S | x_L = b, x_0 = 1) \, \text{Prob}(x_L = b | x_0 =1 ; \ep) \, .
\end{align}
Above, $\text{Prob}(h = t |\ep)$ is the probability that the terminated avalanche has a height $t$. The expression for this is already known from Sec. \ref{hscal}. Let us define the Laplace transform $\ti {\m P}^{\ep} (\lambda, t) :=  \int_0^{\infty} dS \, e^{-\la S} {\m P}^{\ep} (S, t)$. We similarly define Laplace transforms for all other probabilities, and take a Laplace transform of (\ref{sizedisc}). Ultimately we will take the continuum limit, and scale $b,S$ etc. with $L$. It is wise to delay taking this limit as much as possible.\\

Let us first calculate the term $G^{\ep}(b+a,a;t) := \text{Prob}(x_t = b +a | x_0 =a ; \ep)$, for a half-space random walk with the jump probabilities in (\ref{jumprob}). We use the method of images \cite{redner}, which proceeds as follows. We can consider the total contribution of all paths from $a$ to $a+b$ in the full space $\mathbb{Z}$, and then subtract out the contribution of all paths that cross $x = 0$ at least once. Consider such a path $\m C = \{x _j\}$, which crosses $x = 0$ for the first time at $t = t_1$. Now consider a complimentary path $\m C' = \{ x'_{j}  \}$ with $x'_j = x_j \, \forall j \geq t_1$ and $x_l' = - x_l $ for $1 \leq l < t$. $\m C'$ is a path that starts at $-a$ and ends at $a+b$. From (\ref{jumprob}), the ratio of probabilities of the two paths is $\frac{p(\m C')}{p(\m C)} = \left(\frac{1+\ep}{1-\ep} \right)^{2a}$. Clearly the set of paths $\{ \m C\}$ and $\{\m C' \}$ are in one to one correspondence with each other. So, we have :
\begin{align}\label{gg0}
    G^{\ep}(b+a,a;t) = G_0^{\ep}(b+a,a;t) - \left(\frac{1-\ep}{1+\ep} \right)^{2a}G_0^{\ep}(b+a,-a;t) \, .
\end{align}
$G_0^{\ep}(x,y;t)$ is the probability for the walker to travel from $y$ to $x$ with paths in the full space $\mathbb{Z}$. This quantity can easily be found from (\ref{anhw}). Using the result in (\ref{gg0}), we get 
\begin{align}
    G^{\ep}(b+a,a;t) = \int_0^{2\pi} \frac{dq}{2 \pi } \, e^{iqb} \left[ 1-\left( \frac{1-\ep}{1+\ep}\, e^{iq}\right)^{2a}\right] \times \left[ \frac{1-\ep^2}{2} + \frac{1+\ep^2}{2} \cos q -i\ep \sin q\right]^t \, . 
\end{align}
For later purposes it will be useful to take the continuum limit in the previous equation.  Taking $q \rightarrow0$, we get the probability for the continuum case to be:
\begin{align}
    G_c^\ep(b+a,a;t) = \frac{\exp \left( -\frac{(b-\ep t)^2}{t \left( 1-\ep^2\right)}\right)}{\sqrt{\pi t \left(1-\ep^2 \right)}} \times \left[1- \exp \left( -\frac{4a(a+b)}{t(1-\ep^2)}+\frac{4a\ep^3}{1-\ep^2} \right) \right] \, .
\end{align}
Our random walker begins at $x = 1$ on the discrete lattice. In the long time limit we expect the displacements to be of order $b \sim t^{1/2}$, in which case we can take the small $a$ limit in the above equation. \\

Next, it will be useful to calculate certain `path sums', where we perform a weighted sum over directed paths on the discrete lattice. These will later be coarse grained into path integrals. Let $Z_{a,b}^{\ep}(S, t)$ be a weighted sum over all paths $\{x_j\}$ with $x_0 = a, x_t = a+b , \,b \geq 0$, and size $\sum_{j=1}^{t}x_j = S$. 
\begin{align}
    Z_{a,b}^{\ep}(S, t) = \sum'_{\{x_j \}} \omega^{\ep}[\{x_j \}] \times \delta _{S, \sum_jx_j} \Rightarrow \ti Z_{a,b}^{\ep}(\la, t) = \sum'_{\{x_j \}} \w^{\ep}[\{x_j \}] \times \prod_{j}e^{-\la x_j} \, .
\end{align}
Above $\sum'$ is a constrained sum over paths satisfying above initial conditions, and $\w[\{ x_j\}]$ is the weight of the path. 
The weight of each path is defined according to the random walk jump probabilities (\ref{jumprob}). Let $N_{\pm}, N_0$ be the number of forward jumps, backward jumps and `no jumps'. Then $N_0 + N_+ +N_- = t $ and $N_+ - N_- = b$. So we get the Laplace transformed path sum to be :
\begin{align}\label{pathsum}
    Z_{a,b}^{\ep}(\la, t) = \left(1-\ep^2 \right)^t\left( \frac{1+\ep}{1-\ep} \right)^{b}\sum'_{\{x_j \}} 2^{-\left( 2N_++2N_- + N_0\right)}\times \prod_{j}e^{-\la x_j} = \left(1-\ep^2 \right)^t\left( \frac{1+\ep}{1-\ep} \right)^{b} Z_{a,b}^{0}(\la, t)\, .
\end{align}
We have expressed the path sum with biased jump probabilities, in terms of a path sum with unbiased weights. Unbiased path sums have been studied for Bernoulli random walkers in half spaces, which are known as Brownian excursions \cite{darling, takacs, meander}. Firstly, coarse graining the unbiased random walk yields the following equation for the continuum probability density of the random walker : $\p_t \,  p(x,t) = \frac{1}{4}\p_x^2 \rho(x,t) + \m O\left(\p^4\right)$. Following the treatment used in \cite{meander}, we can coarse grain our path sum into a path integral : 
\begin{align}\label{PI}
    Z_{a,b}^{0}(\la, t) &\simeq  \int_{x(0) = a}^{x(t) = a+b} \m D x (t) \, \exp \left( -\int_0^t ds \, \left[\dot x^2 + \la  x(s)+ V_0(x)\right]\right) \, . \\
    V_0 (x) &= \begin{cases}
        0, &x > 0 \\
        \infty, &x < 0 \, . \nonumber
    \end{cases}
\end{align}
Above, $V_0$ is simply a hard wall potential added to constrain the paths to the positive half space. We would like to evaluate $Z_{a,b}^{0}(\la, t)$, as well as the special case $Z_{a,b}^{0}(0, t)$. For these two cases we can write the path integral (\ref{PI}) as simple matrix elements $\bra{a+b}  e^{-Ht} \ket{a}$ and $\bra{a+b}  e^{-H_0t} \ket{a}$. Here, $H$ and $H_0$ are Hamiltonians given by :
\begin{align}
    H_0 = -\frac{1}{4}\p^2 + V_0(x) \, , \, H = H_0 + \lambda x \, .
\end{align}
The eigenfunctions of $H_0$ are $\{ \ket{k} | k \geq 0\}$ with $\braket{x|k} = \sqrt{2/\pi}\sin(kx)$. The eigenvalues are simply $u_k = k^2/4$. The matrix element, therefore, is :
\begin{align}
    Z_{a,b}^{0}(0, t) = \bra{b+a} e^{-H_0t} \ket{a} = \frac{e^{-b^2/t}}{\sqrt{\pi t}} \left[ 1- e^{-4a(a+b)/t}\right] \, .
\end{align}
The normalised eigenfunctions $\{ \ket{n} | n \geq 1\}$  and eigenvalues $\{u_n\}$ of $H$ are given by:
\begin{align}
    \psi_n(x) := \braket{x|n} = (4 \la)^{1/6}\, \frac{\Ai \left[ (4\la)^{1/3}x - \al_n\right]}{\Ai'(-\al_n)} \, , \, u_n = \left(\frac{\la}{2}\right)^{2/3} \al_n \, .
\end{align}
Above, $\Ai(z)$ is the Airy function of the first kind, and $-\al_n$ are the zeros of the Airy function, which satisfy $0 < \al_j < \al_{j+1}, \, \forall j$. Now the matrix element is easily expressed as: 
\begin{align}
    Z_{a,b}^{0}(\la, t) = \sum_{n\geq 1} \psi_n(a) \psi_n(a+b) \, e^{-u_nt} \, .
\end{align}

We are now in a position to calculate $\ti {\m P }^{\ep}(\la,L)$. In (\ref{sizedisc}). taking the large $L$ limit and converting all sums to integrals, we get:
\begin{align}\label{probint}
    \ti {\m P }^{\ep}(\la,L) \simeq   \int_0^{L} dt \, \rho^{\ep}(t) \frac{Z_{1,1}^{\ep}(\la,t)}{Z_{1,1}^{\ep}(0,t)} +\int_0^{\infty} db \, G_c^{\ep}(b+a,a;L) \frac{Z_{1,b}^{\ep}(\la,L)}{Z_{1,b}^{\ep}(0,L)} \equiv I_1 + I_2  \, .
\end{align}
Above, $\rho^{\ep}(t) = \frac{\left( 1-\ep^2\right)^t}{2 \sqrt{\pi}t^{3/2}} $ is the probability distribution of avalanche heights. It is fortunate that $Z_{a,b}^{\ep}(\la,L)/Z_{a,b}^{\ep}(0,L)$ is independent of $\ep$ (\ref{pathsum}). 

We define scaled variables $\xi = \ep L^{1/2}, r = bL^{-1/2}, \tau = tL^{-1}, v = (\la/2)^{2/3}L$. For large $L$, we can make the following substitutions.
\begin{align}
    \rho^{\ep}(\tau) &\rightarrow  L^{-3/2} \, \frac{e^{-\xi^2 \tau}}{2 \sqrt{\pi}\tau^{3/2}} \nonumber \\
    G_c^{\ep}(b+a,a;L) &\rightarrow L^{-1} \, \frac{4r}{\sqrt{\pi}}e^{-(r-\xi)^2} \nonumber \\
    \frac{Z_{1,1}^{\ep}(\la,t)}{Z_{1,1}^{\ep}(0,t)} &\rightarrow 2 \sqrt{\pi} \, (v \tau)^{3/2}\sum_{n>0} e^{-\al_n v\tau} \nonumber \\
    \frac{Z_{1,b}^{\ep}(\la,L)}{Z_{1,b}^{\ep}(0,L)} &\rightarrow \sqrt{\pi} \frac{ve^{r^2}}{r} \sum_{n > 0} e^{-\al_n v } \times \frac{\Ai \left[ 2vr - \al_n \right]}{\Ai'(-\al_n)} \, .    
\end{align}
We use the above to complete the integrals in (\ref{probint}), and obtain $\ti {\m P }^{\ep}(\la,L)$. Let $\chi_n(z) := \int_{-\al_n}^{\infty} dx \, e^{xz} \Ai(x)$. Inverting the Laplace transform yields :
\begin{align}
    \m P^{\ep}(S,L) &= L^{-2} \, {\varphi}\left( \ep L ^{1/2}, \,  2SL^{-3/2} \, \right) \, , \nonumber \\
    \ti \varphi(\xi,v) &\equiv  2 \ti\Omega \left(\xi, v^{2/3} \right) \, , \nonumber \\
    \ti \Omega(\xi,v)  &\equiv \sum_{n>0} \left[ v^{3/2}\,\frac{1-e^{-\xi^2-\al_nv}}{\xi^2 + \al_nv} + \chi_n \left( \frac{\xi}{v} \right) \frac{e^{-\al_n (v - \xi/v)-\xi^2}}{\Ai'(-\al_n)} \right] \, .    
\end{align}
Above, $\ti \varphi, \ti\Omega$ are the Laplace transforms of $\varphi, \Omega$ in the second argument. A special case which will be useful later is the distribution in the absence of an excess density ($\xi \rightarrow 0$). This limit can  easily be taken, and the resulting equations are the same as those in \cite{meander}.

\subsubsection{Form of the SROT curves}\label{sizeroc}

The excess density exhibits Gaussian fluctuations in the large $L$ limit, and the typical excess density (\ref{density}) scales as $\ep \sim L^{-3/4}$. Just as for heights in Sec. \ref{hscal}, the excess density only nontrivially affects the size distribution $\m P^{\ep}(S,L)$ for $\ep \sim L^{-1/2}$. \\

Let $\Xi = \ep L^{-3/4} \equiv \xi L^{-1/4}$. The prediction efficiency index $c$ is defined in the same manner as that for the height case. The parametric expressions for the SROT curves are given by:
\begin{align}\label{rocsize}
    \nu(\Xi) &= \frac{1}{2} \text{erfc} \left( \Xi \sqrt{\frac{m}{2}}\right) \nonumber \\
    c(\Xi,a) &= \frac{\Phi(\Xi,a)}{\Phi(-\infty,a)} \, , \, \Phi(\Xi,a) := \sqrt{\frac{m}{2\pi}}\int_{\Xi}^{\infty} du \, e^{-mu^2/2}\int_a^{\infty}d\mu \, \varphi\left(uL^{-1/4},2 \mu\right)  \, .
\end{align}
For large $L$, we can expand (\ref{rocsize}) in negative powers of $L$. Doing this yields :
\begin{align}
    c-\nu &= \frac{\,e^{-\Xi^2}}{\sqrt{2\pi m }} \times L^{-1/4}g(a)\,; \, g(a) \equiv \frac{\int_{2a}^{\infty} d\mu \, \p_u \varphi(0,\mu)}{\int_{2a}^{\infty} d\mu \, \varphi(0,\mu)} \, .
\end{align}
Above, $\p_u\varphi(u,\mu)$ indicates a partial derivative with respect to the first argument. We see that the peak height in the $c-\nu$ vs $\nu$ curves falls as $L^{-1/4}$. The function $g(a)$ is not singular for $0<a \sim \m O(1) $. So in the thermodynamic limit, the SROT curves are simply $c(\Xi)-\nu(\Xi) = 0$, and the toppling number variables are not predictable. \\

To have the curves for different $L$ and $a$ collapse, the variables $L,a$ must lie on contours with $L^{-1/4} g(a)$ constant. The function $g(a)$ can be found by inverting the Laplace transform in $\ti \varphi(u,v)$. Let $\bt_n := \al_n (6a)^{-2/3}$. Then we get :
\begin{align}
    g(a) &= \frac{3\sum_{n}c_nI_n^{(2)} }{\sum_n \left[ (a\bt_n)^{-1}I_n^{(1)}-d_nI_n^{(0)}\right]} \nonumber \\
    I_{n}^{(0)} &= \int_0^{\bt_n} du \,\kappa(u) \, , \, I_n^{(1)} = \int_{0}^{\bt_n} du \, u \kappa(u)\, , \, I_n^{(2)} = \int_{\bt_n}^{\infty} du \left(u - \bt_n\right) \kappa(u)  \, , \nonumber \\
    \kappa(u) &\equiv \exp \left( -\frac{2}{3}u^3\right) \times \left[ u \Ai \left( u^2 \right) - \Ai' \left( u^2 \right) \,\right] \, , \nonumber \\
    c_n &=\frac{\chi_n'(0)+\al_n \chi_n(0)}{\Ai'(-\al_n)} \, , \, d_n = \frac{3\chi_n(0)}{\Ai'(-\al_n)} \, .
\end{align} 
The function $g(a)$ vanishes for $a\rightarrow 0$. This means that small sized avalanches ($a \sim 0.1$) are almost unpredictable even for finite $L$. 

\subsection{Comparison to results from other studies}\label{compare}
In \cite{shapoval}, the authors study a Manna sandpile of size $L\times L$. For the Manna model, the probability distribution for toppling numbers is of the form :
\begin{align} \label{mann}
    \text{prob}(S) \sim \begin{cases}
        S^{-\bt}, & S \lesssim L^{D}\\
        L^{-\bt D}g_M\left(SL^{-D}\right), &S >L^{D}   \, .     
    \end{cases}
\end{align}
Here $D = 2.75$ is a scaling exponent for sizes, and $g_M$ is a scaling function. It is known that $g_M(u)$ is roughly constant for small $u$, but decays as $\exp \left( -u^{b}\right)$ for large $u$ and some exponent $b$. is The authors find that extremely rare, superscaled avalanches of size  $\sim L^{D'}, D' > D$ are predictable using the excess density in the thermodynamic limit, to the extent that the sum of the false positive and false negative prediction rates asymptotically tend to zero.\\

On the other hand, we have shown that in the thermodynamic limit both height and toppling number variables are unpredictable using the excess density in directed sandpiles. For the directed sandpile, the lack of multiple site topplings imposes a cutoff on on avalanche sizes. Such avalanches are too small to be significantly affected by the typical density fluctuations in the system. But in the Manna model, sites can topple multiple times, and there is a possibility of extremely large superscaled avalanches. Their probability is non-trivially affected by the typical density fluctuations in the system, due to which they can be reliably predicted. However, these avalanches occur with a probability $\sim \exp \left( -L^{b(D'-D)}\right)$, so this asymptotically very high predictability is not really useful in practice. 

\section{Summary}
We have studied the predictability of large avalanche events in the directed sandpile model. Given a time series of avalanche sizes, we have considered prediction strategies depending on a single variable constructed from past data. For the first strategy, this variable is the waiting time since the previous large event. Global events in general are found to be unpredictable. Local events show better predictability as seen from their SROT curves. We find exact expressions for these SROT curves in terms of the waiting time distribution for large events (Figs. \ref{local L 256 var A}, \ref{local L 400 var A}). The predictability with this strategy is shown to be scale invariant, and thus persists in the thermodynamic limit. \\

A second class of strategies use the excess density in the sand pile to make predictions about large events for the next time-step. For both height and toppling number variables, the exact expressions for the SROT curves have been found. For this strategy, we have shown using both analytically and numerically that predictability only exists for finite sizes, and decays with increasing lattice sizes. This conclusion is restricted to directed sandpiles, and notably different from Manna sandpiles, where extremely large events show good predictability in the thermodynamic limit. \\

\section{Acknowledgements}

I thank Prof. Deepak Dhar for much needed guidance and many useful discussions, as well as a critical review of this manuscript. Many thanks to the International Centre for Theoretical Sciences, Bengaluru, where much of this work was completed. I acknowledge the support of the NIUS programme of HBCSE-TIFR funded by the Department of Atomic Energy, Govt. of India (Project No. 2024-T15).

\bibliography{sand}

\end{document}